\documentclass[12pt]{iopart}

\begin{document}

\title[Indirect methods]{Coulomb dissociation, 
a tool for nuclear astrophysics}

\author{G Baur}

\address{Forschungszentrum J\"ulich, Institut
f\"ur Kernphysik, D-52425 J\"ulich, Germany}
\ead{g.baur@fz-juelich.de}

\author{S Typel}

\address{GANIL, Bd. Henri Becquerel, BP 55027, F-14076 Caen
Cedex 05, France}
\ead{typel@ganil.fr}

\begin{abstract}
 A short status report on Coulomb dissociation, an indirect method
for nuclear astrophysics is given.
An analytically solvable approach to study electromagnetic
excitation in $^{11}$Be, the archetype of a halo nucleus,
is proposed.

\end{abstract}

%\maketitle

\section{Introduction}
%v1.tex (+v1save) in npa3 directory
Cross sections for nuclear reactions of astrophysical interest can be 
very small, too small to be directly accessible experimentally.
Sometimes indirect methods can be useful
to determine astrophysical S-factors. The ANC (Asymptotic Normalization 
Constant) method \cite{akram}, the Trojan horse method
\cite{stefano, marco,annphys, enam} and 
the Coulomb dissociation method \cite{stefan}
were discussed at the present conference.
A more recent minireview is given in \cite{mumbai}.
In all these methods, some theoretical interpretation 
of the experimental results is necessary in order to
arrive at an astrophysical S-factor. 

In the Coulomb dissociation method a fast projectile passes
through the Coulomb field of a heavy nucleus. The time-varying
electromagnetic field experienced by the projectile acts 
like a spectrum of equivalent photons which can dissociate the 
projectile. 
The equivalent photon spectrum can be computed from the 
kinematics of the process.
In this way one can determine photodissociation cross-sections,
essentially based on quantum electrodynamics.
Radiative capture cross sections leading to the ground state 
of the final nucleus (the projectile) can be deduced by detailed 
balance. The projectile can be an exotic nucleus. Thus one has the 
unique possibility to study the interaction of unstable nuclei
with photons \cite{sonntag,konstanze}.
In particle physics the Primakoff effect \cite{prima}
has been used for some time to study the interaction of 
photons, pions, $\Lambda '$s, etc.\ with photons.  
Another method to study the interaction of photons with 
exotic nuclei are e-A colliders \cite{carlos}. 
Coulomb dissociation is the only method for
many years to come.

The theoretical description of intermediate energy Coulomb 
excitation and dissociation
has been developed over the past decades, it is reviewed in 
\cite{bht}. The main characteristics are given in Sect.~2.
We propose as a `homework problem' a quasirealistic simple
model, which can be solved analytically. It could also serve 
as a benchmark for tests of more sophisticated or involved
models like the Continuum Discretized Coupled Channels model
(CDCC). 
A major result of radioactive
beam physics  is the discovery of low-lying electric dipole strength
in neutron-rich nuclei, see, e.g.,\cite{aum}. 
This discovery was made possible by the `working horse' Coulomb
dissociation. 
This is discussed in Sect.\ 3.
Low lying dipole strength is  directly related to the halo structure
of these nuclei. The effective-range approach to low lying E1 strength and
simple scaling laws are discussed. 
In some astrophysical scenarios for the r-process
it is vital to know the low lying electric dipole strength.
With the future radioactive beam facilities it will be possible
to access these questions. 

\section{Theory of intermediate energy Coulomb excitation
and dissociation}

One of the basic parameters in Coulomb excitation 
is the ratio of the collision time to
the nuclear excitation time, the so-called adiabaticity parameter
\begin{eqnarray}
\xi=\frac{\omega b}{\gamma v}
\end{eqnarray}
where $\omega$ is the excitation energy, b is the impact
parameter and  v the beam velocity. The corresponding
Lorentz factor $ \gamma$ is typically not much larger than one.
For $\xi \ll 1$ the process is sudden, and excitation is possible; 
for $\xi \gg 1$ the
system follows adiabatically the time varying field and the 
excitation probability tends to zero. The strength of
the excitation is measured by the 
strength parameter
\begin{eqnarray} \label{chidef}
\chi^{(\lambda)}=\frac{Z_T e \langle f|M(E\lambda)|i\rangle}{\hbar 
v b^\lambda} \: .
\end{eqnarray}
The projectile, characterized by the electromagnetic 
matrixelements $\langle f|M(E\lambda)|i \rangle$ 
is excited on its passage through
the Coulomb field of the (heavy) target nucleus with charge $Z_T$.
The parameter $\chi^{(\lambda)}$ can be regarded as the number of exchanged
photons.
In electromagnetic excitation it is a good approximation that
the nuclei do not penetrate each other. In this case, the interaction 
is encoded in  the electromagnetic matrixelement 
$\langle f|M(E\lambda)|i \rangle$ between the
relevant states. The Coulomb parameter $\eta$ 
is the monopole strength parameter, i.e.\ $\lambda=0$ in (\ref{chidef}), 
and the 
multipole matrixelement is replaced by the charge $Ze$ of the nucleus.  
For higher beam velocities  higher order effects tend to be small.
For not too light nuclei $\eta$ is still $\gg 1$, and the semiclassical
description is appropriate. 

\subsection{A quasirealistic and analytically solvable model of Coulomb
excitation of neutron halo nuclei}

An archetype of a halo nucleus is $^{11}$Be with a $^{10}$Be
core and a single halo neutron in the $2s_{1/2}$-state. There is a 
strong E1 transition to the $1/2^-$ bound state,
the only bound excited state of the system.  
This dipole transition  was studied 
by Coulomb excitation 
at GANIL, RIKEN and MSU \cite{Ann95,Nak97,Fau97,Sum07}. 
In Ref. \cite{Sum07} higher order effects are studied in the 
'XCDCC' approach. We take this as an opportunity to revisit 
the theoretical study
of higher order effects in intermediate energy Coulomb 
excitation. 
In \cite{tyba95} electromagnetic excitation of $^{11}$Be  
is studied in the sudden limit of the semiclassical method.  
Higher order effects are treated to all orders.
In \cite{tyba01} an analytically solvable model for higher order 
electromagnetic excitation effects of neutron halo nuclei was presented.
In that work, there was only the s-wave bound state. 
Now we consider the case where there is, in addition, 
a p-wave bound state, as it is the case in $^{11}$Be.
In the sudden approximation the dipole excitation amplitude
is  given by 
\begin{equation} \label{asudden}
a_{\rm sudden}= \langle f|\exp\left(- \rmi \vec q_{\rm Coul} \cdot \vec{r}
\right) |i \rangle
\end{equation}
where $\vec{q}_{\rm Coul} = \frac{2 Z Z_{\rm eff}e^2}{\hbar v b} \vec{e}_x$.
The impact parameter b is chosen to be in the x-direction.
(For the actual calculation in polar coordinates it is convenient to
change this to the z-direction.)
The dipole approximation is quite
well fulfilled, the dipole effective charge $Z_{\rm eff}^{(1)}
=\frac{Z_{c} m_n}{m_n + m_c}$ is much larger than the 
corresponding quadrupole charge. The neutron and core mass
are denoted by $m_n$ and $m_c$ respectively, the charge of the core
is given by $Z_c$.
The sudden approximation
is applicable for $\xi \ll 1$. Even for the comparatively
low  GANIL energies of about 40 MeV/nucleon
this is reasonably well fulfilled.
The most important intermediate states are expected to be
in the low energy continuum, where the dipole strength has a peak,
at around 1 MeV excitation energy.
 The sudden approximation has the advantage 
that intermediate states are treated by closure, thus one only needs a 
model for the initial and final states, and not for all the 
intermediate states. 
In lowest order in $q_{\rm Coul}$ the first order dipole approximation is 
obtained.
It is shown in \cite{tyba95} that third order
E1 excitation is more important than second order E1-E2 excitation.
The matrixelements are dominated by the exterior contributions.
In a pure single particle model the radial wave functions
of the $2s_{1/2}$ and $1p_{1/2}$ states are given by  
\begin{eqnarray}
f_0(r)=C_0 q_{0}r  h_0( \rmi q_{0} r)
\end{eqnarray}
and 
\begin{eqnarray}
f_1(r)=C_1 q_{1}r  h_1( \rmi q_{1} r)
\end{eqnarray}
Both states are halo states and the normalization 
constants are given by $C_0= \sqrt{2q_0}$ and $C_1
=\sqrt{2q_1^2 R/3}$ in the halo limit respectively \cite{tyba05}. 
The bound state parameters $q_i$ ($i=0,1$) are related to
the binding energies by $E_i=\frac{\hbar^2 q_i^2}{2\mu}$
where $\mu$ is the reduced mass of the core-neutron system. 
We have $E_0= 504$~keV and $E_1=184$~keV.
The radius of the core is denoted by R.  
With these model asumptions we can calculate 
the B(E1) value for the $1/2^+ \rightarrow 1/2^-$-transition
as well as the 
higher order effects in electromagnetic excitation. 
Whereas in \cite{Sum07} quite sophisticated models are
used, our approach is simple, transparent and 
at the same time close to reality.
We propose this to be a model study and leave the spectroscopic factors
equal to one. (They could be adjusted, which would result in a quasi-
realistic description of the $^{11}$Be system for our purpose.) 
The XCDCC calculations are quite involved, with many parameters.
It would be very useful to check the method by comparing to a simple
case, such as this one, where analytical results are possible.
In order to avoid nuclear effects a sharp cutoff at a minimum
impact parameter $b_{\rm min}$ can be introduced.

The $B(E1)$ value is given by $ B(E1)=\left(Z_{eff}^{(1)} e\right)^2
|R_{01}|^2/(4\pi)$, where $R_{01}$
is the radial dipole integral. 
The integrals are elementary.
Using $\int_0^\infty r^n e^{-ar}\rmd r= n!/a^{n+1}$
we find 
\begin{equation}
 R_{01}=\frac{2 \gamma_{0}^{1/2}(\gamma_0 + 2\gamma_1)
}{3^{1/2}(\gamma_0 + \gamma_1)^2} R
 \: ,
\end{equation}
where $\gamma_0=q_0R$ and
$\gamma_1=q_1 R$. We note that we extended the 
radial integral over the exterior wave function from R to zero.
For $R \rightarrow 0$ the radial dipole integral goes to zero because the 
normalization of the p-wave function tends to zero in this limit.
Thus R must be kept finite, say $R= 2.78$~fm \cite{tyba04},
this value determines the asymptotic normalization of the 
p-wave bound state. We find  $B(E1)=0.193$~$e^2$fm$^2$,
to be compared to the  value of $B(E1)=0.105(12)$~$e^2$fm$^2$ obtained from
an analysis of the GANIL data, see
\cite{Sum07}, consistent with other Coulomb dissociation 
experiments at RIKEN and MSU
and the value obtained by the Doppler shift attenuation method \cite{mill}.

We expand the excitation amplitude (\ref{asudden}) in terms of 
the dimensionless strength parameter $y=q_{\rm Coul}/(q_0 + q_1)$.
(Cf.\ (\ref{chidef}), we take $e/(q_0 + q_1)$ as a convenient 
measure for the order of magnitude of the dipole matrix-element.) 
The excitation probability is given by $P(b)=|a_{\rm sudden}|^2$. 
The lowest order term is proportional to $y^2$.
The most important higher order contribution comes from
the third order in $q_{\rm Coul}$. It can be calculated analytically.
Its interference with the lowest order term leads to the next term
in the expansion in $y$, of the order of $y^4$. We have $P(b)=
P_{LO} + P_{NLO} + ...$. 
The lowest order term is given by
\begin{equation}
P_{LO}= y^2 \frac{4 \gamma_0 (\gamma_0 + 2\gamma_1)^2}{27 (\gamma_0
+\gamma_1)^2}\equiv C_2/b^2
\end{equation}
The next term is found to be 
\begin{equation}
P_{NLO}=-y^4\frac{8\gamma_0^{3/2}(\gamma_0 +2 \gamma_1)
(\gamma_0 + 4 \gamma_1)}{45(\gamma_0 + \gamma_1)^2}
\equiv - C_4/b^4
\end{equation}

Total cross sections are obtained by integration over the impact 
parameter, starting from a minimum impact parameter $b_{\rm min}$.
The sudden approximation fails for large impact parameters, and 
an adiabatic cut-off $b_{\rm max}=\gamma v/\omega$ has to be introduced
for the lowest order result. We put $\omega=320$~keV, the 
energy of the $1/2^{-}$ state in $^{11}$Be. (For the higher order terms
this is not necessary, the convergence in b is fast enough.)
We get 
\begin{equation}
\sigma_{LO}=2\pi C_2 \ln \frac{b_{\rm max}}{b_{\rm min}}
\end{equation}
and
\begin{equation}
\sigma_{NLO}= - \frac{\pi C_4}{b_{\rm min}^2} \: .
\end{equation} 
We note that the strength parameter $y$ is proportional to $1/v$,
i.e.\ the leading order term decreases like $1/E$, the
next-to-leading order term like 
$1/E^2$, where $E$ is the beam energy.
We think that this analytical model could serve as a benchmark
for tests of more involved reaction models. 
We hope to publish a more detailed account of the present approach in the 
future.

\section{Low-lying electric dipole strength in neutron rich nuclei}

An effective-range approach to low lying E1-strength for
one-neutron halo nuclei was developed in \cite{tyba05,tyba04}.
There is a small parameter 
\begin{equation}
\gamma \equiv qR=\frac{R_{\rm halo}}{R}.
\end{equation}
In lowest order, the dipole strength is independent of $\gamma$.
The $B(E1)$-strength function 
is proportional to the shape function 
$S_{l_i}^{l_f}$ and scales with the parameter $x^2\equiv 
E/E_{\rm bind}=q^2/q_0^2$, where $E$ is the c.m.\  energy
in the continuum. For s-p transitions it is given by \cite{tyba04}
\begin{equation}
S_0^1=\frac{x^3}{(1+x^2)^2}\left[1-a_1 q^3(1+3 x^2)\gamma^3 +...\right] \: .
\end{equation}
This remarkably simple result can be applied to 
deuteron photodisintegration, and $s_{1/2}$-neutron halo nuclei
like $^{11}$Be, $^{15}$C, $^{19}$C,...
The interaction of the final state p-wave neutron
with the core can usually be neglected.
Thus low lying strength due to transition to a 
structureless continuum is found. 
It may look like a resonance, but it has nothing to do with a resonance.
This was recognized long time ago \cite{hansen, beba88}. 
We quote from a recent review of low lying dipole strength
\cite{dario}: '...the onset of dipole strength in the low-energy region
is caused by nonresonant independent single-particle
excitations of the last bound neutrons'.
In general there are characteristic effects 
of the core-neutron interaction in the continuum state,
usually more pronounced for states with $l_f=l_i-1$.
For the s-p transitions this term is proportional
to $\gamma ^3$, which is quite small for a halo nucleus.
For the low energies relevant here this
interaction can be parametrized in terms of the scattering length.
An interesting effect of this type was found by analysing the 
high precision data of $^{11}$Be Coulomb dissociation \cite{palit}.
A large scattering length $a^{j=1/2}_{l=1}= 456$~fm$^3$ 
was found \cite{tyba04}.
It is due to the $p_{1/2}$-subthreshold state.  
A treatment of two-neutron halo nuclei in the effective
range method for low lying strength of halo nuclei is given
in \cite{erice}.

\section{Conclusion and Outlook}

Electromagnetic excitation is a powerful tool to investigate
the interaction of (quasireal) photons with unstable nuclei.
It will continue to play a prominent role at the future radioactive 
beam facilities.
%............referenz zu deinen rechnungen?!!......
 A good theoretical understanding of the process
and its interplay with nuclear excitation is mandatory, see, e.g.,
\cite{stefan}.
In the future rp-process nuclei will come into focus.
The possibility of 2p-capture is also discussed. It will
never be possible to study this process in the laboratory.
However, the time-reversed process of Coulomb dissociation with
two protons in the final state is well within reach. 
An example is  Coulomb dissociation of 
$^{17}$Ne \cite{grigor}, where the soft dipole mode in this proton-rich
nucleus is discussed. By a suitable Coulomb dissociation 
experiment valuable information on the 2p-capture cross section on
$^{15}$O at astrophysical conditions approriate 
for explosive burning in novae and X-ray bursts may be obtained.

At future radioactive beam facilities  r-process nuclei will become available.
In certain scenarios (see, e.g., \cite{dario}) it will be important to know 
the low lying E1 strength, which will decisively influence
the r-process abundances.

\end{document}